\documentclass[12pt,twocolumn]{aastex631}
\begin{document}

\title{The Decline and Fall of ROME. V. A Preliminary Search for Star-Disrupted Planet Interactions and Coronal Activity at 5 GHz Among White Dwarfs within 25 pc}

\author[0000-0001-6987-6527]{Matthew Route}
\affiliation{Department of Physics and Astronomy, Purdue University, 525 Northwestern Avenue, West Lafayette, IN 47907, USA}

\correspondingauthor{Matthew Route}
\email{mroute@purdue.edu}


\begin{abstract}

The recent discovery of planetesimals orbiting white dwarfs has renewed interest in the final chapters of the evolution of planetary systems.  Although observational and theoretical studies have examined the dynamical evolution of these systems, studies of their magnetic star-planet interactions, as powered by unipolar induction, have thus far only been assessed theoretically.  This fifth paper of the ROME (Radio Observations of Magnetized Exoplanets) series presents the results of a targeted mini-survey of nine white dwarfs within 25 pc without known stellar mass companions in search of radio emissions generated by magnetic interactions between white dwarfs and their planetary remnants.  This $\sim$5 GHz Arecibo radio telescope survey achieved mJy-level sensitivity over $<$1 s integration times.  Although no exoplanet-induced stellar radio flares were detected, this is the first survey to search for magnetic star-planet interactions between white dwarfs and planetary companions, cores, or disrupted planetesimals.  It is also the most extensive and sensitive radio survey for intrinsic coronal emissions from apparently isolated white dwarfs.  The study of radio emissions from white dwarf systems may present a new means to detect and measure DC and DQ magnetic fields, search for white dwarf coronae, characterize the density and spatial distribution of white dwarf magnetospheric plasma, characterize the dynamical and electrical properties of planetary cores, and offer new constraints on the modeling of double degenerate merger events.

\end{abstract}
 
\section{Introduction}

An emerging focus of white dwarf astrophysics is the discovery and characterization of planetary systems, or their remnants, about them.  Approximately 25\% of white dwarfs show evidence for recent accretion of metal-rich planetary debris given their atmospheric ``pollution,'' while in several cases, tidally disrupted planetesimals or remnant planetary cores have been identified that are currently contributing to this process \citep{kaw16,van15,man19,van21}.  The evidence to support these discoveries includes metal absorption lines that ``pollute'' their stellar spectra despite the $\lesssim$ million-year timescales over which heavier elements should settle in their interiors \citep{koe09}, the detection of infrared excesses in their spectral energy distributions \citep{bar16,xu20}, and in rare instances, the presence of metallic emission lines and periodic decreases in optical flux that are unrelated to the host white dwarf rotation periods.  The origin of these systems is currently debated, although the migration of individual planets likely results from a combination of stellar mass loss, tidal drag, bow shock drag during temporary engulfment, and Lorentz drift \citep{li98,wil05,ver19}.  This latter effect stems from Lorentz torques that dissipate some of the system gravitational potential energy as heat within the interior of the planetary body, and some as the kinetic energy of the planet.  Lorentz torques are created when an orbiting planetesimal encounters a uniform, but time-varying white dwarf magnetic field.  A unipolar induction circuit is then established between the white dwarf atmosphere and the ionosphere or conducting surface of the planetary remnant, which acts as the electromotive force (EMF).  Thus, the magnetic interaction between a white dwarf and an orbiting planetesimal/planetary core may be a critical ingredient in determining the origin and rate of evolution of such ``polluted'' systems \citep{li98,kaw14,bro19,ver19}.

Among main sequence stars and potentially ultracool dwarfs at the bottom of the main sequence, star-planet magnetic interactions hold the promise of elucidating the magnetic characteristics of orbiting exoplanets (see \citealt{rou23} and references therein).  The magnetism in such systems is thought to be revealed by photospheric spots, chromospheric Ca II H\&K and H$\alpha$ activity, and coronal X-ray and radio emissions (e.g., \citealt{rou19}).  Although white dwarfs are thought to lack many of the structures of main sequence stars, theoretical work suggests that unipolar induction should generate detectable non-thermal electron cyclotron maser (ECM) emissions at radio wavelengths, just as it may for main sequence star-planet interactions \citep{ww04,wil05}.  Moreover, \citet{wil04} argued that the X-ray emission in the ultrashort ($P<$10 min) white dwarf binary RX J0806+15 and the anti-phased X-ray versus optical/infrared emission in RX J1914+24 indicate the presence of unipolar induction interactions.  Although this model has also been proposed to explain the antiphased light curve vs. H$\alpha$/H$\beta$ emission in WD 1639+537 (GD 356), problems with this explanation have recently come to light \citep{ver19,wal21}.

The template for the study of unipolar induction is the Jupiter-Io system and its legion of resulting  electromagnetic signatures.  In this system, Io (the EMF) traverses the Jovian magnetosphere, and a circuit is established connecting Io's surface to the Jovian ionosphere via a magnetic flux tube \citep{gol69}.  Electrons traversing the circuit spiral around magnetic field lines that converge at Jupiter's magnetic poles.  This convergence of field lines coupled with the existence of field aligned potential drops can create loss-cone or shell velocity distributions that enable the emission to occur as the local cyclotron frequency, $\nu_{c}$, exceeds the local plasma frequency \citep{tre06}.  The maximum frequency of ECM radio emission is given by $\nu_{c}$[MHz] $= 2.8 n B$ [Gauss], where $n$ is the harmonic number, and $B$ is the local magnetic field strength.  The observable hallmarks of ECM emission include $\sim$100\% circular polarization fractions and brightness temperatures $T_{B}\gtrsim$10$^{10}$ K \citep{dul87}.  Within the Jovian system, ECM emissions occur at $\nu\lesssim$40 MHz, corresponding to the maximum magnetic field strength in Jupiter's polar regions.  At increasing radii above the poles, electrons encounter a rapidly diminishing global dipolar magnetic field, resulting in ever lower frequency radio emissions \citep{lad94}.  Electrons that enter the atmosphere collide with and excite atoms and molecules, that then emit in the ultraviolet at the auroral footprints \citep{dep15}.

Although theory suggests that single white dwarfs hosting planetary systems and double degenerate systems may be sources of radio emissions, many systems containing a white dwarf companion are known radio emitters.  For instance, \citet{bar20} determined that 73\% of cataclysmic variables (CVs) are sources of ECM emission, although in these cases, the emission characteristics point to an origin in the companion M dwarf magnetosphere.  Within ``white dwarf pulsar'' systems, another exciting developing vein of research in white dwarf astrophysics, it is unclear whether radio emissions originate within the M dwarf magnetosphere, or are governed by the magnetic interaction of the white dwarf and M dwarf magnetospheres as in the case of AR Sco \citep{mar16,buc17,lyu20}.

Given these trends in research, it is surprising that but a single reference to a search for radio emissions from isolated white dwarfs has been published, yielding a non-detection \citep{fer97}.  Progress in observational discovery, theory, and technology suggest reasons to reopen this avenue of research.  In addition to the impressive observational discoveries concerning white dwarf pollution, white dwarf pulsars, and radio loud CVs, $\gtrsim$20\% of white dwarfs are known to be magnetic with fields $B\gtrsim$40 kG \citep{bag21}.  Thus, magnetic activity among white dwarfs is little-explored, although the particle acceleration that results from magnetic reconnection may be amenable to radio detection.  Should ECM radio emissions be detected from white dwarf magnetospheres, they may enable the precise measurement of magnetic fields from DC white dwarfs, where nearly featureless spectra preclude the precision measurement of their fields $B\lesssim$100 kG, and warm DQ white dwarfs, where thus far, few are known to be magnetic presumably due to the low signal-to-noise ratios of their spectra \citep{fer20,ber23}.

DA white dwarfs with effective temperatures $T_{eff}\lesssim$18,000 K and DB white dwarfs with $T_{eff}\lesssim$50,000 K begin to develop convective envelopes. All DC and DZ white dwarfs have convective envelopes \citep{sau22}.  Convection may generate magnetic fields that drive starspots, as confirmed by studies of WD 1953-011, and potential quasi-coronal activity, the constraints on which are rather weak (e.g, \citealt{max00,wei07,bri13}). On the sun, ECM coronal radio emissions above sunspots have been detected, illustrating a link between the two phenomena \citep{yu24}.  Thus, in addition to measurement of global magnetic fields, smaller, regional fields may also be detectable and measurable by their related radio emissions.

Another reason to search for radio emissions from white dwarfs without interacting stellar companions is that radio technology has improved significantly since the non-detection of GD 356 \citep{fer97}.  This can best be exemplified by comparing the performance parameters of a survey for magnetospheric exoplanet emission at the time of that review with recent work.  For instance, radio observations of the warm Jupiter system 70 Vir reached a 3$\times$ rms noise sensitivity of 1.8 mJy at 1.4 GHz after 10 s integrations across a 50 MHz bandpass \citep{bas00}, as compared to the 1.0 mJy sensitivity at 4.75 GHz across a $\sim$1 GHz bandpass in 0.9 s achieved by my ROME IV survey of the same target \citep{rou24}.  Thus, recent technology yields greater-than-order-of-magnitude increases in sensitivity and bandpass coverage, which enables a greater range of magnetic fields to be probed, than decades ago.

Intrigued by these opportunities, I conducted a mini-survey of nearby white dwarfs without known stellar companions at radio wavelengths.  Section 2 describes survey targets and observations, while  Section 3 presents the non-detection results.  Section 4 discusses the implications of this survey, and the promising role that radio observations may play in characterizing white dwarf interactions with planetary remnants, coronal activity, magnetism in DC and DQ white dwarfs, and plasma properties in their magnetospheres.  Section 5 concludes by describing future research directions that can build and improve upon this work.

\section{Target Selection and Observations}
White dwarfs within 25 pc of the Sun were selected using the pre-{\em Gaia} catalog of \citet{sio09}, since this project was conceived in the fall of 2016 (Table 1).  This distance threshold was established to maximize the chance of detecting even the weakest ECM emissions for a given instrumental sensitivity.  Targets were also selected to be without known close companions to interact with, such as M dwarfs in CVs.  Although the {\it Spitzer Space Telescope} InfraRed Array Camera (IRAC) and Wide-field Infrared Survey Explorer (WISE) indicate that several selected targets have infrared excesses, subsequent investigations did not find evidence of debris disks around them \citep{bar16,xu20}.  Given the single, fixed-dish nature of the Arecibo radio telescope, targets were also required to have declinations of 0$^\circ$ to +38$^\circ$.

Targets were observed with the 305 m William E. Gordon radio telescope at Arecibo Observatory (AO) from 2017 February 9 to 2017 April 18 under observing program A3124 (PI: Route).  Nearly all systems were observed for 1--3 hr, sometimes across multiple epochs (Table 2).  These observing epochs consisted of 600 s on-target science scans bracketed by 20-second calibration on-off scans that use a local oscillator.

Radio emissions were acquired by the C-band receiver which has a center frequency of 4.75 GHz.  The C-band receiver has a 5'' pointing accuracy and an ellipsoidal half-power beam width of $\sim$1', where the beam semimajor axis is in the zenith direction.  The antenna gain and system temperatures vary with zenith angle from 5.5--9.0 K Jy$^{-1}$ and 25--32 K, respectively.  Signals received by the dual-linear polarization C-band receiver were processed by an array of seven field-programmable gate array (FPGA)- equipped Fast Fourier Transform (FFT) Mock spectrometers (Chris Salter, personal communication).  Each 8192-channel spectrometer has a 172 MHz bandpass that can be individually tuned such that it overlaps that of its neighbors by 30 MHz, yielding an aggregate bandpass of $\sim$1 GHz from 4.239 to 5.262 GHz.  Natively-developed software computed full Stokes parameters, performed flux calibration, bandpass correction, and iterative radio frequency interference (RFI) removal \citep{rouphd}.  Although data were sampled at a (frequency, time) resolution of (20.9 kHz, 0.1 s), they were rebinned to (83.6 kHz, 0.9 s) resolution to enhance their signal-to-noise characteristics.

Afterwards, for each science scan, dynamic spectra and bandpass-averaged time series graphs of flux density were created for every Stokes parameter for every spectrometer.  Since this program primarily sought ECM emission (from star-planet interactions) or mildly-relativistic gyrosynchrotron emission (potentially from magnetic reconnection intrinsic to white dwarf coronae), flares of $\gtrsim$10\% circularly polarized (Stokes V) emission were sought (e.g., Figure 1).  Additionally, evidence for relativistic synchrotron flares was sought in Stokes Q and U.  Pulsars are known to emit $\sim$100\% linearly polarized bursts at radio wavelengths (e.g., \citealt{rad69}), and similarly, \citet{buc17} found that optical emissions from AR Sco were up to 40\% linearly polarized.  On the other hand, 1.5-9 GHz radio observations of AR Sco measured linear polarization fractions of only about 5\% \citep{sta18}, and \citet{bar20} found about 3/4 of their sample of CVs were strongly circularly polarized, but none were linearly polarized.  In either case, candidate bursts were identified by surpassing a 3$\sigma$ noise threshold in their Stokes V (circular) or Stokes QU (linear) polarization time series graphs and had their morphology compared with those found in a library of RFI artifacts accumulated over multiple AO survey efforts \citep{rou13,rou16b,rou23,rou24}.

\section{Results}

No circularly or linearly polarized radio bursts were detected from any white dwarf target system.  Since AO is insensitive to quiescent (slowly varying and/or unpolarized) radio emission due to its local calibration procedure and confusion limitations, only strongly polarized radio bursts of several minutes' duration or less could be detected.  I emphasize that this survey was sensitive only to $\gtrsim$10\% circularly polarized bursts of ECM or gyrosynchrotron radio emission, or linearly polarized bursts of synchrotron emission.  

Detection limits of the radio emission from target white dwarfs can be computed using 3$\times$ the standard deviation ($\sigma$) of the frequency-integrated time series from the Mock spectrometer most devoid of RFI, with a frequency range of 4.522--4.694 GHz (box 2).  Table 3 reports the detection limits in both circular and linear polarization.  When flares of circularly polarized emission are detected, they can be verified by lacking simultaneous emission in Stokes QU, which would denote RFI.  However, the reverse is not true as RFI features detected at AO overwhelmingly consist of linear polarization components alone. The detection limits in linear polarization are significantly higher than for circular polarization given the greater challenge of automatically detecting and removing more extensive RFI.  The best and worst detection limits for this survey are depicted in Figure 2 with respect to the radio luminosities of several classes of stars and compact objects.  These detection limits constrain the luminosity of stellar flares induced by magnetic star-planet interactions, in addition to those that might be initiated by intrinsic coronal activity powered by dynamo processes \citep{ise17,sch21}.  One noteworthy feature of this survey relative to previous efforts \citep{rou13,rou16b} is the much-worsened RFI environment at AO in 2017 relative to 2008--2010.  Every data set was too corrupted by RFI from 5.090-5.262 GHz (box 6) to be useful, as were nearly all data sets from 4.948--5.120 GHz (box 5).

The observing epoch durations can be compared to the white dwarf rotation and putative exoplanet orbital periods to determine their respective phase coverage fractions.  Among our targets, only WD 1134+300 has a measured rotation velocity \citep{pil84}, but data quality issues exclude it from further analysis (Table 2).  Thus, target temporal properties will be analyzed in a statistical fashion.  The rotation periods of magnetized white dwarfs have a bimodal distribution, where one group rotates on timescales $\gtrsim$100 yr, and a second group with periods ranging from $\sim$1 hr to days \citep{bri13,kaw15}.  The average rotation period of ZZ Ceti stars observed by {\it Kepler} is 35 hr, while the rotation period of the most rapidly rotating isolated white dwarf, SDSS J221141.80+113604.5, is 70 s \citep{her17,kil21b}.  By these standards, the survey's $\sim$2 hr observations yield average rotation phase coverage of $\sim$6\%, but could, optimistically, reach $\sim$100\% or, pessimistically, are negligible for the slowest rotators. Using the description of the Roche limit cast in terms of the minimum orbital period ($P_{min}$) as a function of orbiting body bulk density ($\rho$), $P_{min}[hr]\approx$12.6$\rho^{-1/2}$[g cm$^{-3}$], and assuming a Fe/Ni planetesimal core ($\rho$=8 g cm$^{-3}$), $P_{min}$= 4.45 hr.  Thus, the average orbital phase coverage is $\Delta\phi_{orb}\lesssim$45\% for conducting planetary cores on the verge of tidal disruption \citep{bro19,van21}.

\section{Discussion}

\subsection{Implications of Non-Detections for Star-Planet Interactions}
As demonstrated by Figure 2, this survey could detect $\gtrsim$10\% circularly polarized flares, and so would have been sensitive to a wide range of ECM and gyrosynchrotron flare luminosities.  In accordance with the star-planet unipolar dynamo interaction model constructed by \citet{ww04}, this 5 GHz survey could detect ECM flares from Earth-sized exoplanets in $\lesssim$10 hr orbits, with thermal plasma densities and temperatures of $n_{th}\sim$10$^{8}$ cm$^{-3}$ and $k_{B}T_{th}\sim$1--10 eV, respectively, about a $0.7$ $M_{\odot}$ white dwarf with magnetic moment $\mu=10^{31}$ G cm$^{3}$.  Additional model parameters of the ECM emission can be found in that text, which likely represents the best-case scenario.  ECM plasma and loss cone parameters that differ by a factor of $\gtrsim$5 can make radio emission at this low of frequency undetectable.  RFI degrades the sensitivity to linearly-polarized synchrotron flares, but those powered by rotating white dwarf magnetospheres such as AR Sco would have been observable.

The cyclotron formula in Section 1 shows that the 4.239 GHz to 5.262 GHz frequency range of this survey corresponds to ECM radio emission at the fundamental frequency within $B\sim$1.5 to 1.9 kG magnetic fields.  Such a frequency range corresponds to very weak magnetic fields compared to the  mean surface fields measured on white dwarfs thus far by spectropolarimetry ($B\gtrsim$40 kG). Yet this range may be appropriate for either white dwarfs with extremely weak surface fields, or for emissions originating within certain regions of white dwarf magnetospheres (i.e., plasma at $\sim$8 $R_{WD}$ from a white dwarf with a 1 MG dipolar surface magnetic field would yield emissions detectable by this survey).  \citet{wil05} suggested that planets in $P\sim$10$^{4}$ to 10$^{5}$ s orbits would be most readily detectable at $\nu=$100 GHz, and objects at shorter orbital periods accessible at $\nu\lesssim$650 GHz.  Although more compact systems may produce greater flux densities at high frequencies, the probability of detecting them is low.  This is because decreasing the planet/planetesimal orbital period increases the EMF powering the emission, but decreases the planetary system synchronization timescale.  Modeling indicates that even planets $\sim$tens of Roche radii away from white dwarfs with the strongest fields ($B\sim$10$^{9}$ G) rarely survived for 10$^{8}$ yr \citep{ver19}.  \citet{wil05} estimated that, due to the difficulty of forming compact white dwarf planetary systems, the probability of detecting radio emissions from one was $\sim$0.1\%, yielding approximately 10 detections at $\sim$10 GHz with VLA and $\sim$200 detections at $\sim$250 GHz with ALMA.  Subsequent work indicating that the incidence of white dwarfs hosting planetary systems may be $\sim$25\% likely points to a significantly higher detection probability (e.g., \citealt{kaw16}).

On the other hand, when evaluating the possibility of unipolar induction occurring within the magnetosphere of GD 356, \citet{wal21} listed several potential barriers to its operation.  These include: the potential lack of emitting plasma, the possibility that certain ranges of white dwarf magnetic moment and companion orbital parameters may result in faster-than-light Alfv\'{e}n velocities, and the requirement that significant particle acceleration occur within the flux tubes connecting the magnetized white dwarf with its planetary companion.  There may be at least three sources of plasma within a white dwarf magnetosphere: accretion from remnant particles and collisionally or tidally disrupted planetesimals (e.g., \citealt{far17,bag24a,bag24b}), Bondi-Hoyle accretion (e.g., \citealt{fer97}), and radiation pressure acting on cyclotron resonances (e.g., \citealt{zhe91}). Within the solar system, magnetized objects that exhibit ECM emission have magnetic field-aligned electric fields that accelerate electrons and rarefy the emitting plasma enabling ECM to operate \citep{tre06}.  Thus, several solutions to these objections may exist, but the detection of ECM emission within a white dwarf planetary system would be strong support in favor of unipolar induction.

As described in Section 3, the maximum white dwarf rotation (planetary remnant orbital) phase coverage for surveyed targets is 100\% (45\%).  The detection of ECM radio emission is also complicated by its beaming pattern, with emission more strongly beamed at higher electron kinetic energies.  ECM opening angles range from 90$^\circ$ to 10$^\circ$ with respect to the local magnetic field, with beamwidths of order a few degrees and duty cycles of a few percent \citep{ww04}.  Thus, more promising results may be obtained by observing a larger sample of targets over multiple, longer epochs.

The recent discovery that WD 1632+177 may consist of a double degenerate system separated by 0.03 AU \citep{kil21a} may indicate that the system is too compact to generate ECM radio emissions via unipolar induction from interacting planets/planetesimals (which may orbit one or both components).  On the other hand, the separation between the degenerate components is also too great to support ECM generation via unipolar induction between a magnetized and unmagnetized white dwarf pair operating analogously to the Jupiter-Io system.  Although in the latter case, this radio survey was conducted at the optimal frequency of 5 GHz for typical white dwarf electromagnetic characteristics \citep{ww04}.

\subsection{Implications of Non-detections on Intrinsic Coronal Activity}
Thus far, this paper has focused on radio emissions from white dwarfs instigated by their planetary systems.  However, a constraint on the radio emissions from white dwarfs also serves as a constraint on their \emph{intrinsic} coronal emissions.  The hot, rarefied plasma of a stellar corona can be diagnosed by their X-ray and radio emissions (e.g., \citealt{ben10}).  In particular, the empirical G\"{u}del-Benz relationship expresses a link between the non-thermal particle acceleration observable at radio wavelengths within solar and stellar flares and the X-ray emissions that occur when the accelerated particles thermalize plasma in the lower corona and chromosphere.  As indicated earlier, this radio survey enables a search for white dwarf coronae at lower magnetic field strengths than usually detected via Zeeman spectroscopy and continuum circular polarization (e.g., \citealt{fer20,ber23}), and lower plasma densities than those probed by X-rays, which have thus far found no evidence for coronal activity in, for instance, GD 356 ($L_{X}<$6.0$\times$10$^{25}$ erg s$^{-1}$) (e.g., \citealt{wei07} and references therein).  Using the empirical G\"{u}del-Benz relationship, the radio emission detection limits presented here are equivalent to an X-ray luminosity constraint $L_{X}<$1.2$\times$10$^{29}$ erg s$^{-1}$ for gyrosynchrotron flaring.  Figure 2 also suggests that sensitivity increases of $\gtrsim$10$^{1}$ would be required to detect non-coherent, gyrosynchrotron radio emissions generated by flares within a scaled-down magnetosphere that resembles the Sun in its activity.

A major issue in white dwarf astrophysics is whether they can support coronae and coronal activity at any level.  Although theoretical calculations indicate that $\sim$10$^{10}$--$\sim$10$^{11}$ erg cm$^{-2}$ s$^{-1}$ of wave energy flux may reach a white dwarf's surface, some work suggests that radiative damping and wave trapping of purely acoustic modes within a convective zone may preclude coronal heating \citep{mus87,win94,ulm03,mus05}.  However, weakly magnetic white dwarfs may be able to support coronae through the propagation of traverse slow magnetic wave modes from the convective zone to the photosphere.  It is important in this scenario that the white dwarf magnetic field is weak enough so that convection is not suppressed \citep{mus87}.  Thus, the ability to detect weak magnetic fields that may enable coronal activity will place valuable constraints on this astrophysical process in white dwarfs.  Moreover, whether coronae are detected on white dwarfs may present an unconventional means to test theories that aim to resolve the solar corona heating problem (e.g., \citealt{dem15,kli15}).

\subsection{Characterization of White Dwarf Magnetospheres with ECM Emissions}
The detection of ECM radio emissions from white dwarfs may provide a complementary means to measure the magnetism of white dwarfs from classes that are particularly difficult to characterize via spectropolarimetry.  For instance, since DC white dwarfs lack strong spectral lines, their magnetism cannot generally be measured via the Zeeman effect. However, continuum circular polarization has recently enabled the measurement of $\gtrsim$100 kG longitudinal magnetic fields of $\sim$10 DC white dwarfs (e.g., \citealt{ber22,ber23}).  Similarly, the low signal-to-noise ratios achieved in warm (10,000$<T_{eff}<$16,000 K) DQ white dwarf spectra have made measurement of their magnetic fields difficult \citep{fer20}.  The detection of ECM radio emissions from these two classes of white dwarfs may therefore provide a more precise ($\sim$100 G) measurement of their magnetic field strengths and topologies.  Obviously, in the absence of a detection of isolated white dwarf radio emissions and measurement of their underlying magnetic fields, these opportunities remain speculative.  For the foreseeable future, DC and DQ white dwarf magnetism will be better studied via established techniques such as continuum circular polarization and spectropolarimetry, respectively.

The detection of ECM radio emission from white dwarf planetary systems would provide valuable constraints on the plasma characteristics of white dwarf magnetospheres, regardless of whether the source of such plasma is the white dwarf itself, plasma accreted from disrupted planetesimals, or the interstellar medium.  It is still unclear whether white dwarf coronae exist, although radio emissions may provide a means to detect them at lower energies.  For ECM to operate and be detectable by a distant observer, $\nu_{pl}^{2}/\nu^{2}<<1$, where $\nu_{pl}=e\sqrt{n_{e}/m_{e}\epsilon_{0}}/2\pi\approx 9\sqrt{n_{e}}$ kHz is the electron plasma frequency, and $\nu$ is the cyclotron frequency defined in Section 1.  The helicity of any detected signal, coupled with measurement of the thermal plasma temperature will constrain whether the ordinary ({\it O}-mode) or extraordinary ({\it X}-mode) dominates, thereby producing additional constraints on the plasma density \citep{ww04,wil05,tre06}.  Furthermore, the detection of ECM radio emissions also offers the possibility of diagnosing magnetospheric accretion disk truncation \citep{met12}, the distribution of obscuring dust (e.g., WD 1145+017, \citealt{van15,far17l} ), and cyclotron cooling of the accretion column (although this later effect may require $\sim$nJy sensitivity, see \citealt{far17}.)
	
The detection of ECM radio emissions also presents the opportunity to detect conducting planetary cores approaching their host white dwarfs before tidal disruption reveals them to spectropolarimetry.  The detection and characterization of the radio emissions from such ``declining'' planetary cores may characterize Lorentz drift, planetary core dynamic viscosity, white dwarf electrical conductivity, and provide an additional measure of planetary core mass \citep{ver19}.  ECM radio waves emitted during the tidal disruption and subsequent accretion process may then provide important clues regarding whether dynamo activity is generated by closely orbiting exoplanets, or remaining planetary cores engulfed during the common envelope phase \citep{far17}.  
	
In addition to modeling the generation of ECM emission by unipolar induction from planets and planetary remnants, \citet{ww04,wil04} also suggested that the same type of interaction would be present between a magnetized and non-magnetized white dwarf pair.  Thus, such emissions would provide valuable constraints on the magnetism and plasma environments of white dwarfs in close binary systems, enabling improved modeling of system evolution, and perhaps even leading to the creation of SN Ia events (e.g., \citealt{sch21}).

\section{Conclusion}

This fifth paper of the ROME (Radio Observations of Magnetized Exoplanets) series has explored a possible means by which the magnetic properties of conducting planets and disrupted planetary cores may be detectable via unipolar induction.  Exoplanets and disrupted planetary cores are hypothesized to generate electron cyclotron maser (ECM) radio emissions powered by this mechanism as occurs in the Jupiter-Io system.  Nine apparently isolated white dwarfs within 25 pc were surveyed using the Arecibo radio telescope at $\sim$5 GHz across a $\sim$1 GHz bandpass, thereby achieving mJy-level sensitivity over $<$1 s integration times.  Although dynamic spectra were analyzed to search for ECM, gyrosynchrotron, and synchrotron flaring that may result from unipolar induction or intrinsic coronal activity, no such flares were found, with a 3$\sigma$ sensitivity of $\nu~L_{\nu}\geq 1.059\times$10$^{24}$ erg s$^{-1}$.

Nevertheless, further survey work in this vein should continue.  Although DC and warm DQ white dwarfs are difficult to characterize via broadband circular polarimetry and spectropolarimetry, respectively, their magnetic fields may be amenable to measurement via ECM emission.  Radio emissions may reveal large and small-scale white dwarf magnetic fields and coronae.  The temporal, frequency, and helical properties of ECM emissions may constrain white dwarf magnetospheric plasma densities and spatial distributions \citep{ww04,wil05,tre06,met12,van15,far17l,far17}.  ECM emissions generated by unipolar induction may reveal inward migrating conducting planetary cores before tidal disruption reveals them to photometry or spectropolarimetry.  This would enable characterization of Lorentz drift, planetary core dynamic viscosity, white dwarf electrical conductivity, and planetary core mass \citep{ver19}.  Radio emissions may even provide insight into double degenerate evolution, potentially leading to the formation of SN Ia.
	
ALMA and VLA provide existing capabilities with which radio emissions from white dwarfs, whether isolated or in planetary systems, may be observed, while the future ngVLA\footnote{``ngVLA Performance Estimates,'' https://ngvla.nrao.edu/page/performance} and SKA1-Mid\footnote{``SKA Telescope Specifications,'' https://www.skao.int/en/science-users/118/ska-telescope-specifications.} will be able to search for such signals with increased sensitivity.  Future surveys should target white dwarfs with ``polluted'' atmospheres (i.e. spectral types DAZ and DZ) and those with large relative abundances of heavier metals that sink on shorter timescales and are thus indicative of recent accretion, such as Fe (e.g., \citealt{koe09,bag24a,bag24b}).  Several targets particularly ripe for further investigation include GD 356, due to the presence of H Balmer line emission \citep{wei07}, and WD 0816-310 and WD 2138–332, due to their correlated and periodic variations in magnetic field and metal line strength, indicating that they recently accreted from disrupted planetesimals \citep{bag24a,bag24b}.  Recent accretion would indicate that plasma may still reside in the white dwarf magnetosphere that can generate detectable ECM radio emissions. Future surveys should also target white dwarfs with a range of magnetic field strengths.  White dwarfs with weaker global magnetic fields are promising targets since they lack strong magnetic fields that would suppress the transfer of magnetic wave energy from the interior into coronae \citep{mus87}.   Orbiting conducting planetary cores would also survive for longer lifetimes around such white dwarfs \citep{ver19}.  Alternatively, high-field magnetic white dwarfs present larger magnetospheric cross-sections for interaction with the interstellar medium to supply plasma for intrinsic coronal activity, and would generate ECM radio emissions at higher flux densities \citep{fer97,wil05}.
 
\section{Acknowledgments}

The author thanks Tom Ayres, Stefano Bagnulo, and Steve Kawaler for discussions that have aided the development of this manuscript.  Data storage and analysis support has been made possible with the Theodore Dunham, Jr. Grants for Research in Astronomy.  Support for the presentation of these results (abstract 407.04) at the 56th Annual Meeting of the AAS Division for Planetary Sciences was made possible by the Purdue College of Science Professional Development Fund.  At the time of the observations that are the subject of this publication, the Arecibo Observatory was operated by SRI International under a cooperative agreement with the National Science Foundation (AST-1100968), and in alliance with Ana G. M\'{e}ndez-Universidad Metropolitana, and the Universities Space Research Association.  This research has made use of NASA's Astrophysics Data System and the SIMBAD database, CDS, Strasbourg Astronomical Observatory, France.

\facility{Arecibo}
\software{GSFC IDL Astronomy User's Library, MATLAB}

\clearpage

\begin{deluxetable}{llllllllll}
\tabletypesize{\scriptsize}
\tablecolumns{10}
\tablewidth{0pt}
\tablecaption{White Dwarf Target Properties}
\tablehead{
	\colhead{Name}&
	\colhead{Alternate}&
	\colhead{R.A.}&
	\colhead{Dec.}&
	\colhead{Spectral} &
	\colhead{Distance}&
	\colhead{Magnetic}&
	\colhead{Companionship?}&
	\colhead{Disk?}&
	\colhead{References}\\
	\colhead{(WD)}&
	\colhead{Name}&
	\colhead{({hh}\phn{mm}\phn{ss})}&
	\colhead{(\phn{\arcdeg}~\phn{\arcmin}~\phn{\arcsec})}&
	\colhead{Type}&
	\colhead{(pc)}&
	\colhead{Field (kG)\tablenotemark{a}}&
	\colhead{}&
	\colhead{}&
	\colhead{}
}
\startdata
1134+300 & GD 140 & 11 37 05.10 & +29 47 58.29 & DA2.2 & 15.69 & 6.5654$\pm$3.7108 & None\tablenotemark{b} & No & {\bf 1};{\em 2}; \underline{3}; 4,5,6\\
1257+037 & Wolf 457 & 13 00 09.06 & +03 28 41.06 & DAQZ8 & 16.50 & 11.20$\pm$18.0 & None & & {\bf {\em 7}}; \underline{3}\\
1334+039 & Wolf 489 & 13 36 31.85 & +03 40 45.94 & DC9 & 8.35 & 330$\pm$730 & None & & {\bf 8}; {\em 9}; \underline{3}\\
1344+106 & EGGR 360 & 13 47 24.24 & +10 21 36.00 & DA7.1 & 20.81 & $<$10($\ast$) & None & No & {\bf 1};{\em 10}; \underline{3}; 4\\
1345+238 & EGGR 438 & 13 48 03.01 & +23 34 46.44 & DC9 & 11.86 &  & M5\tablenotemark{c} &  & {\bf 11}; \underline{12}\\
1609+135 & EGGR 117 & 16 11 25.61 & +13 22 17.94 & DA5.4 & 22.27 &  & None & No & {\bf 1}; \underline{3}; 4\\
1632+177 & PG 1632+177 & 16 34 41.85 & +17 36 34.09 & DA4.9 & 25.62 & 0.215$\pm$0.235 & WD\tablenotemark{d} & No & {\bf 1}; {\em 13}; \underline{14}; 4,5\\
1655+215 & EGGR 197 & 16 57 09.86 & +21 26 48.66 & DA5.4 & 20.99 & 0.44$\pm$0.67 & None & No & {\bf 1}; {\em 13}; \underline{3}; 4\\
1705+030 & EGGR 494 & 17 08 07.97 & +02 57 37.08 & DZ7 & 17.85 & $<$15($\ast$)\tablenotemark{e} & None\tablenotemark{f} & No & {\bf 8}; {\em 15}; \underline{16}; 16\\
1814+134 & LSPM J1817+1328 & 18 17 06.50 & +13 28 24.99 & DA10 & 15.13 & $\sigma_{z}\approx $10\tablenotemark{g}  & None &  & {\bf 17}; {\em 18}; \underline{3}\\
\enddata
\tablecomments{Reference numerals with {\bf bold}, {\em italicized}, and \underline{underlined} formatting denote spectral type, magnetic field strength, and companionship references, respectively. The final number(s) in normal font provide disk reference(s).  If no entry is made in the disk column, the target had no indications of infrared excess that led to follow-up work, while ''No'' denotes follow-up observations did not indicate the white dwarf was surrounded by a debris disk.  Uncertainties are provided with magnetic field strength measurements to indicate data quality, and are expressed with the precision found in the respective texts.  All J2000 RAs, DECs, and distances are from \citet{gai21}.  {\bf References.} (1) \citet{gia11}; (2) \citet{byc09}; (3) \citet{too17}; (4) \citet{bar16}; (5) \citet{xu20}; (6) \citet{mor21}; (7) \citet{put97}; (8) \citet{sio90}; (9) \citet{ang81}; (10) \citet{kaw19}; (11) \citet{dup94}; (12) \citet{far05}; (13) \citet{bag18}; (14) \citet{kil21a}; (15) \citet{bag19}; (16) \citet{far09}; (17) \citet{lep03}; (18) \citet{bag21}.}
\tablenotetext{a}{Longitudinal magnetic field measurement $\langle B_{Z}\rangle$; ($\ast$) is appended when disk averaged magnetic field $\langle B_{\ast}\rangle$ is meant instead.}
\tablenotetext{b}{In addition to studies of infrared excess, \citet{mor21} reported the null-detection of transiting planetestimals around the target with CHEOPS at optical wavelengths.}
\tablenotetext{c}{A revised Gaia distance to a binary system with components separated by 198.5'' yields a semimajor axis of 2354 au, as opposed to 2394 au as reported in \citet{far05}.}
\tablenotetext{d}{High resolution spectroscopy indicates a He+CO binary white dwarf system with components separated by 0.0307 au.}
\tablenotetext{e}{Detection limit computed as 3$\times$ measurement uncertainty on the William Herschel Telescope ISIS instrument.}
\tablenotetext{f}{Any potential companion mass constrained to be $<$12 M$_{J}$ from {\em Spitzer} IRAC photometry.}
\tablenotetext{g}{No $\langle B_{Z}\rangle$ detected, with uncertainty reported for a single measurement.}
\end{deluxetable}
\clearpage

\begin{deluxetable}{lccccc}
	\tabletypesize{\scriptsize}
	\tablecolumns{6}
	\tablewidth{0pt}
	\tablecaption{Observations List}
	\tablehead{
		\colhead{Designation}&
		\colhead{Date}&
		\colhead{Scan Range\tablenotemark{$\ast$}}&
		\colhead{Number} &
		\colhead{Time on} &
		\colhead{Total Time on}\\
		\colhead{}&
		\colhead{(yyyy mm dd)}&
		\colhead{}&
		\colhead{of Scans}&
		\colhead{Target (hrs)}&
		\colhead{Target (hrs)}
	}
	\startdata
	1134+300 & 2017 05 18 & 00000\tablenotemark{$\dag$} & 1.0 & 0.17 & 0.17\\
	1257+037 & 2017 02 09 & 00000--01700 & 5.9 & 0.98 & \\
	$\phantom{1257+037}$ & 2017 02 10 & 00000--01700 & 6.0 & 1.00 & 1.98\\
	1334+039 & 2017 02 18 & 00000--02900 & 10.0 & 1.67 & 1.67\\
	1344+106 & 2017 02 09 & 01800--02700 & 3.5 & 0.58 & \\
	$\phantom{1344+106}$ & 2017 02 10 & 01800--03300 & 5.5 & 0.92 & \\
	$\phantom{1344+106}$ & 2017 02 18 & 03000--03200 & 1.0 & 0.17 & 1.67\\
	1345+238 & 2017 02 26 & 00000--02900 & 10.0 & 1.67 & 1.67\\
	1609+135 & 2017 03 25 & 00000--02700 & 9.5 & 1.58 & \\
	$\phantom{1609+135}$ & 2017 04 17 & 00000--02600 & 9.0 & 1.50 & 3.08\\
	1632+177 & 2017 03 25 & 02800--03400 & 2.2 & 0.37 & \\
	$\phantom{1632+177}$ & 2017 03 26 & 00000--03300 & 11.8 & 1.97 & 2.33\\
	1655+215 & 2017 03 26 & 03400--03700 & 1.9 & 0.32 & \\
	$\phantom{1655+215}$ & 2017 04 10 & 00000--02900 & 10.0 & 1.67 & 1.98\\
	1705+030 & 2017 04 17 & 02700--03300 & 3.0 & 0.50 & \\
	$\phantom{1705+030}$ & 2017 04 18 & 00000--02900 & 9.5 & 1.58 & 2.08\\
	1814+134 & 2017 04 12 & 00000--02200 & 8.0 & 1.33 & 1.33\\
	\enddata
	\tablecomments{Characteristics of AO data sets acquired and analyzed to search for magnetospheric radio emissions.  Data sets referenced in this table may be requested by submitting a ticket with the category ``Arecibo Data'' to the Arecibo Observatory Tape Library hosted by Texas Advanced Computing Center (https://www.tacc.utexas.edu/about/help/).}
	\tablenotetext{\ast}{The scan range column denotes the continuous sequence of on-source, calibration-on, and calibration-off scans that focused on the listed target on a given day.  For example, file a3124.20170209.b2s1g0.01800.fits provides the on-source data from the second, cleanest Mock spectrometer (b2) during the observation of WD 1344+106 (scan 01800) that occurred on 2017 Feb 9.  The dynamic spectra and time series graph derived from this data set are shown in Figure 1.}
	\tablenotemark{\dag}{A system malfunction prevented the acquisition of calibration scans for WD 1134+300. Thus, observations of this target are excluded from further analysis and discussion within the text.}
\end{deluxetable}

\begin{deluxetable}{lllll}
\tabletypesize{\scriptsize}
\tablecolumns{5}
\tablewidth{0pt}
\tablecaption{White Dwarf Survey Detection Limits}
\tablehead{
	\colhead{White Dwarf}&
	\colhead{Flux Density Limit}&
	\colhead{Flux Density Limit}&
	\colhead{Circular $\nu$L$_{\nu}$}&
	\colhead{Linear $\nu$L$_{\nu}$}\\
	\colhead{}&
	\colhead{Circular (mJy)}&
	\colhead{Linear (mJy)}&
	\colhead{(erg s$^{-1}$)}&
	\colhead{(erg s$^{-1}$)}
}
\startdata
1257+037 & $<$2.656 & $<$6.777 & $<$3.989$\times$10$^{24}$ & $<$1.018$\times$10$^{25}$\\
1334+039 & $<$2.754 & $<$7.899 & $<$1.059$\times$10$^{24}$ & $<$3.037$\times$10$^{24}$\\
1344+106 & $<$1.337 & $<$4.846 & $<$3.193$\times$10$^{24}$ & $<$1.157$\times$10$^{25}$\\
1345+238 & $<$1.804 & $<$5.407 & $<$1.399$\times$10$^{24}$ & $<$4.194$\times$10$^{24}$\\
1609+135 & $<$1.477 & $<$5.754 & $<$4.039$\times$10$^{24}$ & $<$1.573$\times$10$^{25}$\\
1632+177 & $<$1.505 & $<$7.797 & $<$5.449$\times$10$^{24}$ & $<$2.822$\times$10$^{25}$\\
1655+215 & $<$1.431 & $<$5.114 & $<$3.478$\times$10$^{24}$ & $<$1.243$\times$10$^{25}$\\
1705+030 & $<$4.432 & $<$13.915 & $<$7.784$\times$10$^{24}$ & $<$2.444$\times$10$^{25}$\\
1814+134 & $<$1.626 & $<$5.922 & $<$2.052$\times$10$^{24}$ & $<$7.472$\times$10$^{24}$\\
\enddata
\tablecomments{The second and third columns contain the 3$\sigma$ flux density detection limits for the targets.  1$\sigma$ is computed by the standard deviation of the bandpass integrated time series at a center frequency of 4.608 GHz (box 2) for a given polarization.  Circular polarization limits are determined directly by Stokes V, and linear polarization limits are computed as $\sqrt{Q^{2}+U^{2}}$.  Linear polarization detection limits are much worse due to the more challenging RFI environment at those polarizations.}
\end{deluxetable}

\begin{figure}
	\centering
	\includegraphics[trim = 0mm 0mm 0mm 50mm, clip, width=0.8\textwidth,angle=0]{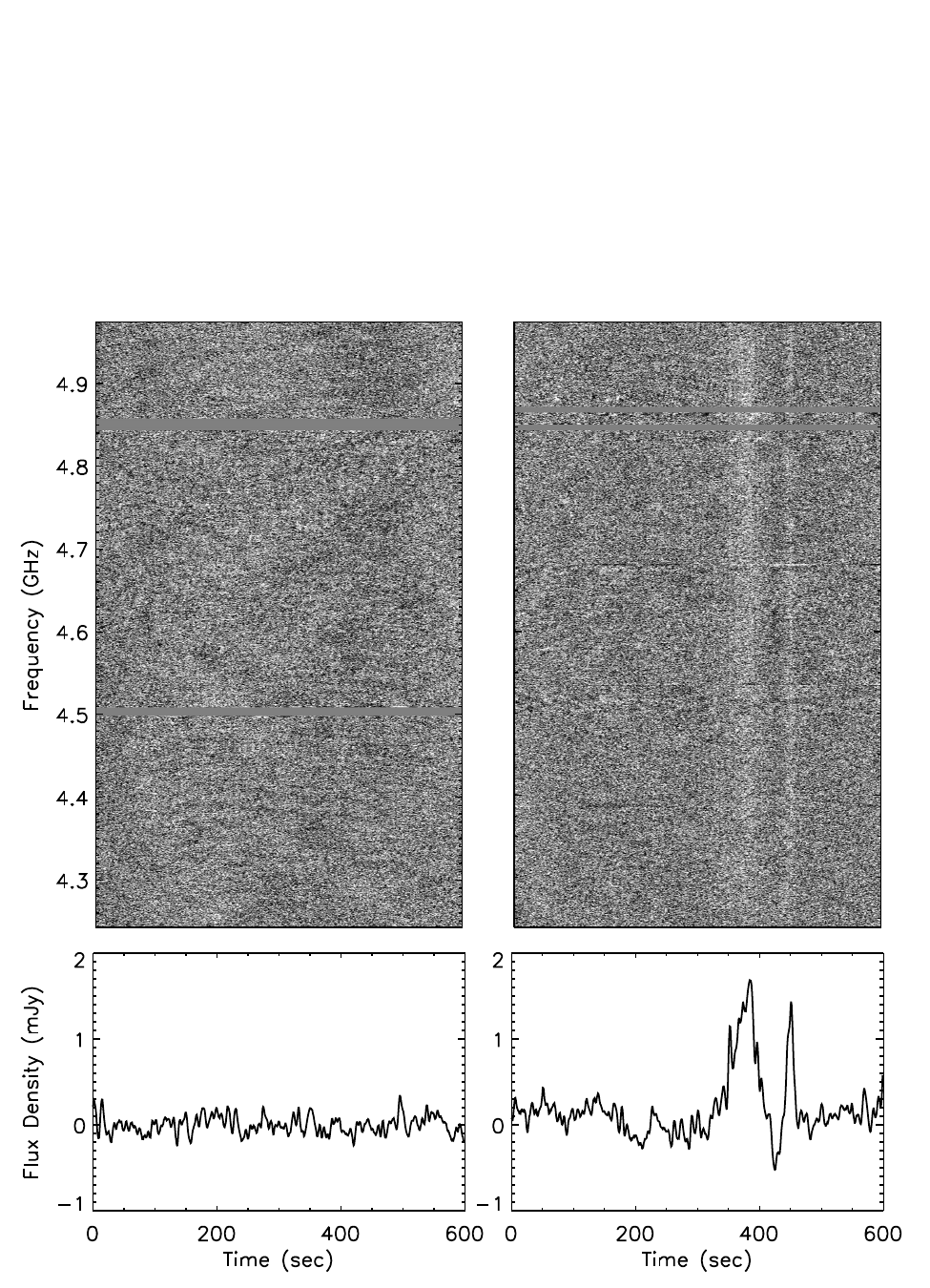}
	\caption{Sample Stokes V dynamic spectra and bandpass-integrated time series from this survey (left) and our previous survey of UCDs (right) at AO.  The left panels exemplify the null-detection of circularly polarized radio emission from WD 1344+106. This data set is the least corrupted by RFI and instrumental artifacts obtained during our survey.  For comparison, the right panels depict a highly circularly polarized ECM radio flare from the T6-dwarf, WISEPC J112254.73+255021.5 \citep{rou16a,rou16b}.  Light (dark) gray denotes left (right) circularly polarized emission. Solid horizontal gray bars near 4.5 and 4.85 GHz mark where strong RFI was excised.  The time series graphs have been smoothed over 3.6 s for clarity.}
\end{figure}

\begin{figure}
	\centering
	\includegraphics[trim = 20mm 0mm 20mm 0mm, clip,width=0.9\textwidth,angle=0]{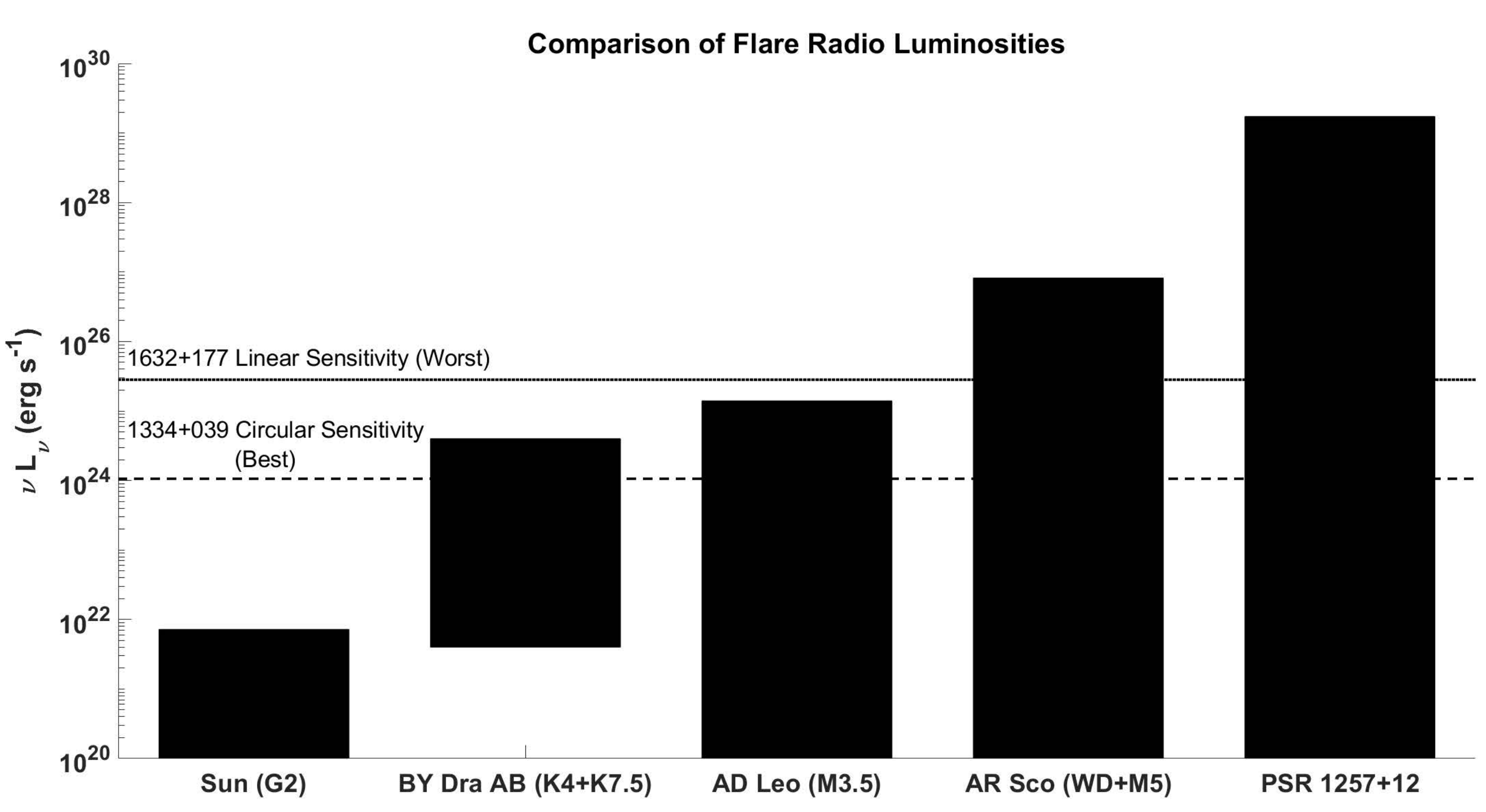}
	\caption{White dwarf survey $\nu L_{\nu}$ radio luminosity detection limits compared to known stellar flare luminosities.  The spectral type(s) of each system are given in parentheses.  The flare radio luminosity of the interacting binary BY Dra AB is calculated from its X-ray flare luminosity ranges via the G\"{u}del-Benz relationship, which relates the X-ray luminosity to radio luminosity of incoherent gyrosynchrotron stellar flares \citep{ben10}.  Beamed flares observed at AD Leo are created by the coherent ECM process \citep{ste01}.  The peak radio emission from AR Sco at 5 GHz is generated by the interaction of the white dwarf's magnetic field with the M dwarf companion's stellar wind, and its radio luminosity determined using the {\em} Gaia measured-parallax \citep{sta18,gai21}.  The luminosity for the planet-hosting pulsar system PSR1257+12 is calculated from its 430 MHz flux density, with distance obtained from Very Long Baseline Interferometry \citep{wol92,yan13}.  The dashed (dotted) lines mark the detection thresholds of our most (least) sensitive observations, those of WD 1334+039 in circular (WD 1632+177 in linear) polarization, relative to these flare energies.}
\end{figure}

\end{document}